\documentclass{iaus}
\usepackage{graphicx}

\title[Simulating AGN feedback] 
{AGN feedback from Jet-ISM/IGM interactions}

\author[Antonuccio-Delogu \& Silk]   
{V. Antonuccio-Delogu$^{1,2}$%
\and J. Silk$^1$}

\affiliation{$^1$Department of Physics, University of Oxford, Keble Road, OX1 3RH, Oxford, United Kingdom \break email: (van,silk)@astro.ox.ac.uk\\[\affilskip]
$^2$INAF, Osservatorio Astrofisico di Catania\break Via S. Sofia 78, I-95123 Catania, ITALY}

\pubyear{2007}
\volume{245}  
\pagerange{1--3}
\date{?? and in revised form ??}
\jname{Formation and Evolution of Galaxy Bulges}
\editors{M. Bureau, L. Athanassoula \& B. Barbuy, eds.}
\begin{document}

\maketitle

\begin{abstract}
We study the propagation of relativistic jets originating from AGNs
within the Interstellar/Intergalactic Medium of their host
galaxies, and use it to build a
model for the suppression of stellar formation within the expanding
cocoon.
\keywords{galaxies: jets, galaxies: bulges, galaxies: ISM}
\end{abstract}

\firstsection 
\section{Jet and surrounding cocoon }\label{sec:jetcocoon}
Relativistic jets from Supermassive Black Holes hosted within the
bulges of spirals and nuclei of early-type galaxies inject
significant amounts of energy into the medium within which they propagate,
\begin{figure}
\centering
 \includegraphics[scale=0.6,angle=90]{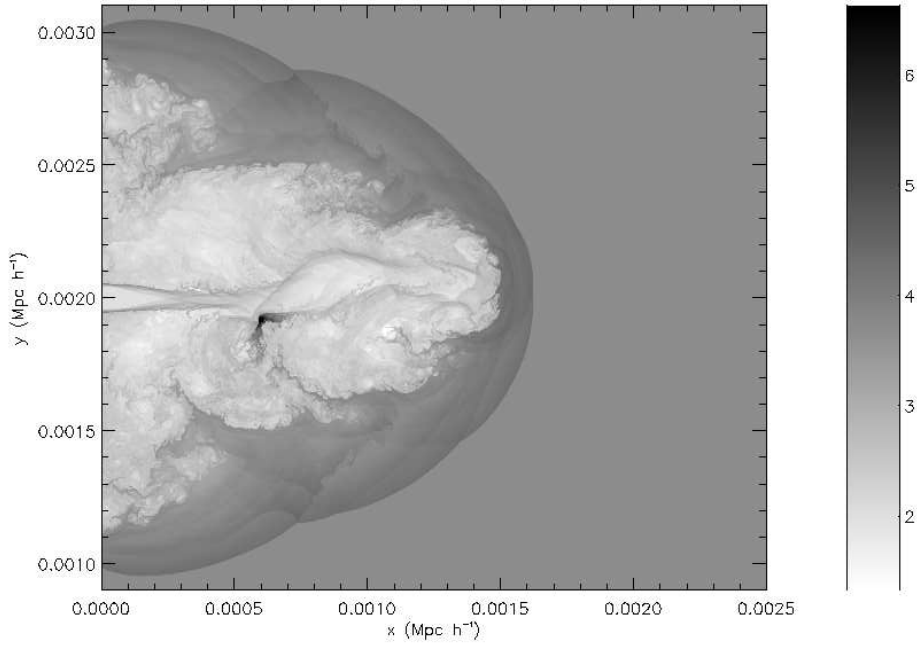}
  \caption{Expansion of the cocoon within the ISM. The jet has an
  input mechanical power $\textrm{P}_{jet} = 10^{46}
  \textrm{ergs}\cdot\textrm{cm}^{-3}\cdot\textrm{sec}^{-1}$, and the ISM
  density is: $\, n_{e}^{ism} = 1 e^{-} cm^{-3}$.}\label{fig:cocoon}
\end{figure}
creating an extended, underdense and hot cocoon. We have performed a
series of simulations of these jets and cocoons using an AMR
code, FLASH 2.5: a typical output is shown in fig.~\ref{fig:cocoon}.\\
We find that a slight modification of the exact models of \cite{1991MNRAS.250..581F} and \cite{1997MNRAS.286..215K}, approximates well the simulations. 

\section{A model for negative AGN feedback}\label{sec:negfeedb}
We can use the exact solutions to develop a model of
\textit{negative feedback} of the AGN on the host galaxy. Within the
cocoon star formation is strongly suppressed, and at any time we
\begin{figure}
 \includegraphics[scale=0.3]{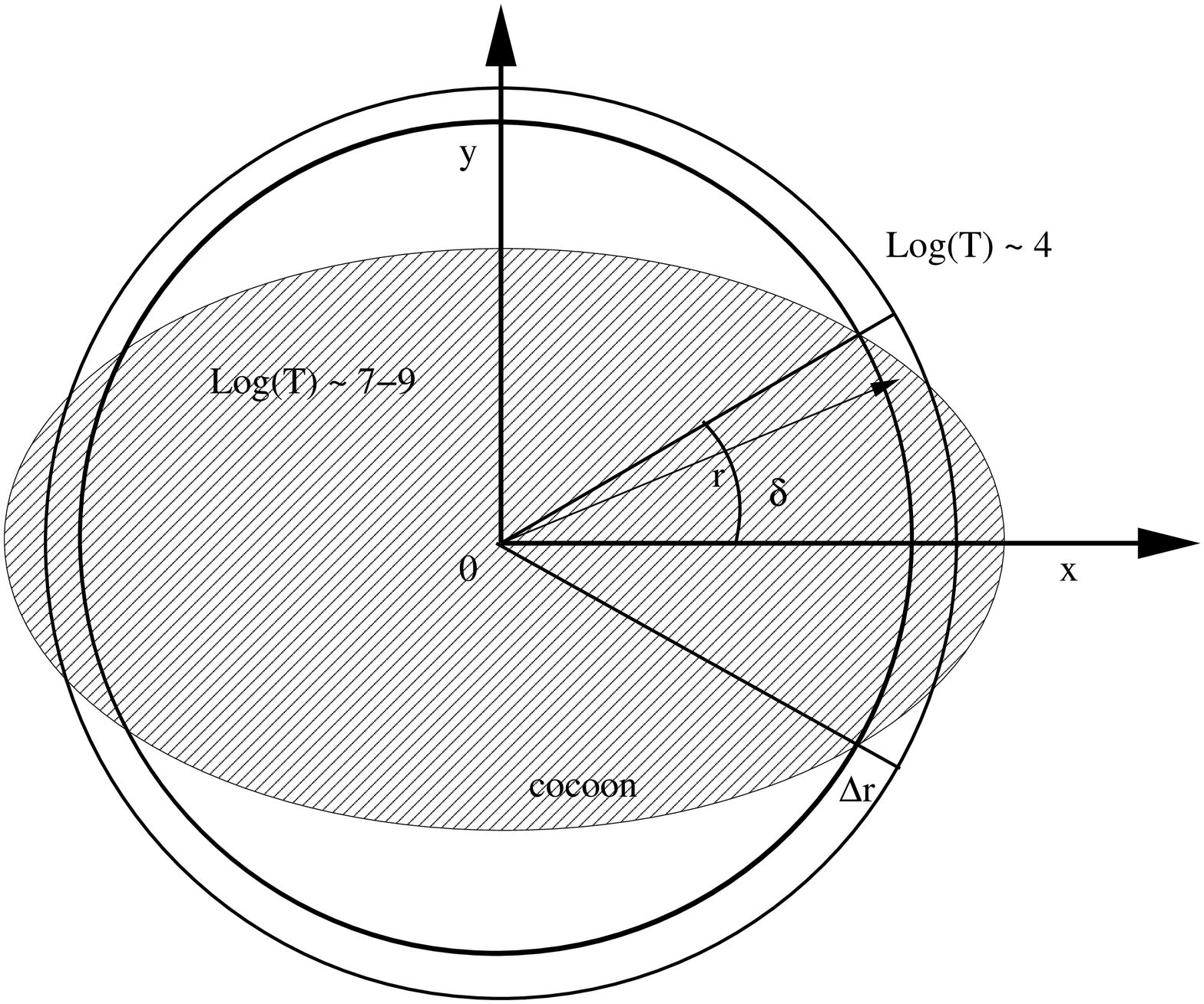}
 \includegraphics[scale=0.3]{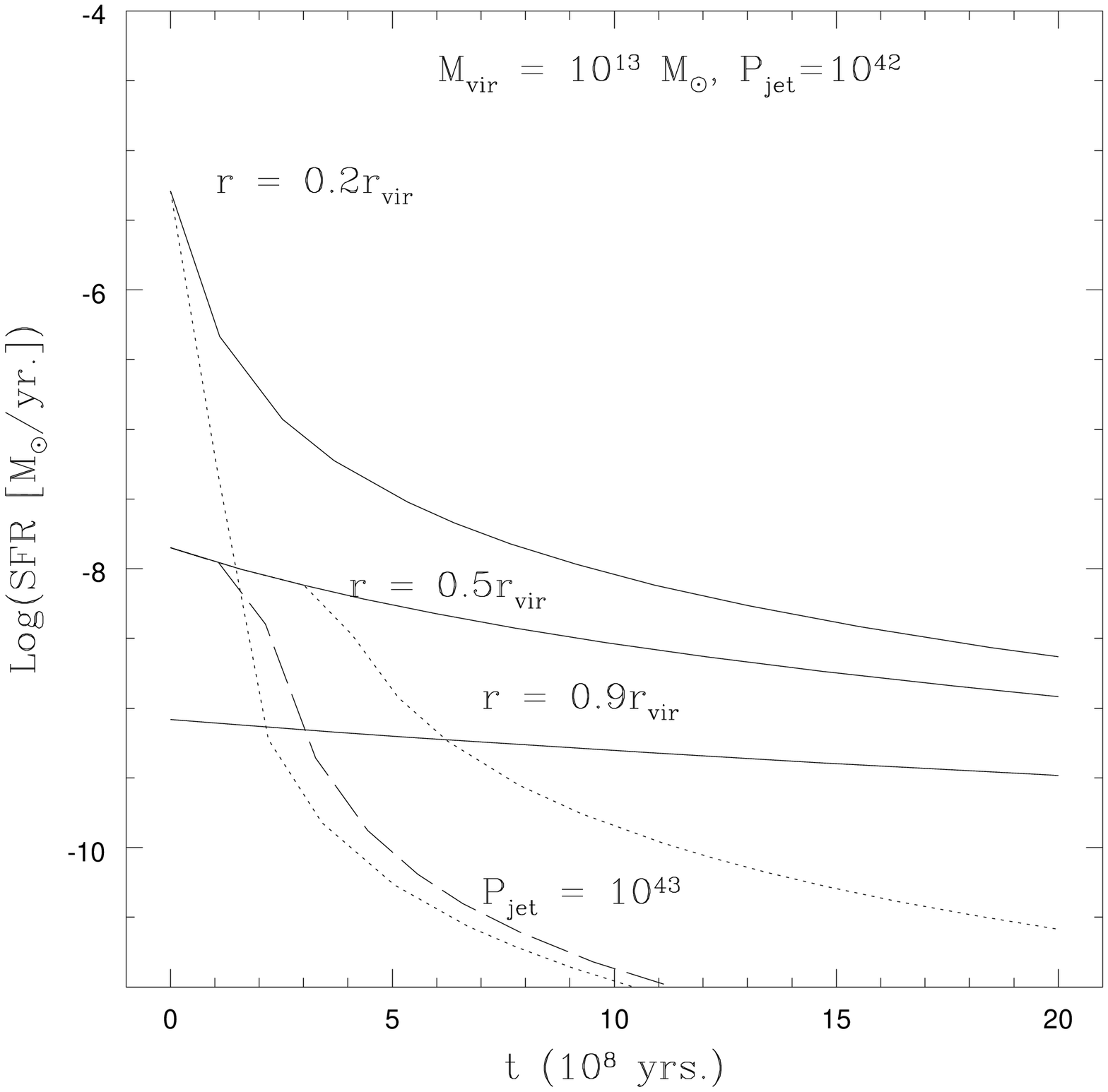}
  \caption{(\textit{Left}: Model for AGN feedback. The shaded ellipsoidal region
  is the expanding cocoon. Stellar formation is suppressed within
  that region of spherical shell of radius $r$ which intersects the
  cocoon.\textit{Right}: Suppression of the SFR, for a M=$10^{11}$
  M$_{\odot}$ host halo.}\label{fig:model}
\end{figure}
know the extent of the region occupied by the cocoon (fig.~\ref{fig:model}, \textit{left}). We show in
fig.~\ref{fig:model} (\textit{right}) the results of this model for AGN feedback. The rapid decrease of stellar
formation is a consequence of the high ISM density in the central
regions, which are the first to be efficiently depleted by the
jet.\\
A more detailed analysis will be presented in a subsequent paper (Antonuccio-Delogu \& Silk, submitted)

\begin{acknowledgments}
V.A.-D. has been supported by EC contract MTKD-CT-002995. FLASH v. 2.5 was partly developed by the
DOE-supported ASC/Alliance Center for Astrophysical Thermonuclear
Flashes at the University of Chicago.
\end{acknowledgments}

\end{document}